\documentclass[english,aps,prc,preprintnumbers,showpacs,superscriptaddress,twocolumn,floatfix]{revtex4-1}
\usepackage[LGR,T1]{fontenc}
\usepackage[latin9]{inputenc}
\usepackage{verbatim}
\usepackage{amstext}
\usepackage{graphicx}

\makeatletter

\DeclareRobustCommand{\greektext}{%
  \fontencoding{LGR}\selectfont\def\encodingdefault{LGR}}
\DeclareRobustCommand{\textgreek}[1]{\leavevmode{\greektext #1}}
\DeclareFontEncoding{LGR}{}{}
\DeclareTextSymbol{\~}{LGR}{126}

\makeatother

\usepackage{babel}
\begin{document}

\preprint{}

\title{Partial conservation of seniority and its unexpected influence on E2 transitions in $g_{9/2}$
nuclei}

\author{Chong \surname{Qi}}

\email{chongq@kth.se}

\selectlanguage{english}%

\affiliation{Department of Physics, KTH Royal Institute of Technology, 10691 Stockholm,
Sweden}

\date{\today}
\begin{abstract}
There exist two uniquely defined $v=4$ states in systems within a
$j=9/2$ subshell, which automatically conserve seniority and do not
mix with other states. Here I show that the partial conservation of seniority plays
an essential role in our understanding of the electric quadrupole
transitions of the semimagic nuclei involving $j=9/2$ subshells, including
the long-lived $8^{+}$ isomer in $^{94}$Ru. The effects of configuration mixing from 
neighboring subshells on the structure of
those unique states are analysed. It is shown that a sharp transition from pure seniority
coupling to a significant mixture between the $v=2$ and $v=4$ states
may be induced by the cross-orbital non-diagonal interaction matrix
elements. Such strong mixture is essential to explain the observed
E2 transition properties of $N=50$ isotones $^{96}$Pd and $^{94}$Ru.
\end{abstract}

\pacs{21.10.Tg, 21.60.Cs,27.60.+j}

\keywords{Insert suggested keywords here --- APS authors don't need to do this.}

\maketitle

One of the greatest challenges in nuclear physics is to understand
the regular and simple patterns that emerge from the complex nuclear
structure. Among those one can mention the shell structure as a consequence
of the strong spin-orbit coupling, which is characterized by nucleons
occupying orbitals with different $lj$ values. While the original
shell model is mostly built upon independent particle motion, the
concept of seniority symmetry has been applied implicitly to account
for the strong pairing correlation. The seniority quantum number refers
to the minimum number of unpaired particles in a single-$j$ shell for a given configuration $|j^n;I\rangle$ 
with total angular momentum $I$ \cite{tal93}. The seniority coupling
has shown remarkable success in describing the spectroscopy and electromagnetic
transition properties of semi-magic nuclei restricted to a single
$j$ shell. Of particular interest are nuclei that can be well approximated
by the seniority coupling in high $j$ orbitals like $0f_{7/2}$.
For heavier systems, we can mention the neutron-rich $^{70-78}$Ni
isotopes \cite{PhysRevC.93.034328}, the $N=50$ and 82 \cite{PhysRevLett.111.152501}
isotones in the $0g_{9/2}$ proton subshell, neutron-rich isotopes
$^{134-140}$Sn with in the $1f{}_{7/2}$ subshell \cite{PhysRevLett.113.132502}
as well as $^{210-218}$Pb in the $1g_{9/2}$ neutron subshell \cite{PhysRevLett.109.162502}. 

Seniority remains a good quantum number within a subshell when $j\leq7/2$.
All states in such systems can be uniquely specified by the total
angular momentum $I$ and seniority $v$. The interaction matrix elements
have to satisfy a number of constraints in order to conserve seniority
when $j>7/2$. For a subshell with $j=9/2$, where all but one two-body
matrix elements conserve seniority, the condition reads \cite{VanIsacker201473,Qi10,tal93,PhysRevLett.87.172501,PhysRevC.67.014303,PhysRevC.82.014304}
\begin{equation}
65V_{2}-315V_{4}+403V_{6}-153V_{8}=0,\label{eq:1}
\end{equation}
where $V_J=\langle j^2;J|\hat{V}|j^2;J\rangle$ denotes a two-body
matrix element and $J$ the angular momentum of a two-particle state $|j^2\rangle$.
The symmetry is broken for most effective interactions (see, e.g., Ref. \cite{Gross76}) in subshells with $j\geq9/2$ where
the eigenstates would be admixtures of states with different seniorities.
For a system with $n=4$ identical fermions in a $j=9/2$ shell, there are three $I=4$ (and also $I=6$) states, which may be constructed so that one state has seniority $v=2$ and the other two have seniority $v=4$. In principle, those seniority $v=4$ states are not uniquely defined and any linear combination of them would result in a new set of $v=4$ states.
However, it was noticed that in the $j=9/2$ shell two special $v=4$
states with $I=4$ and 6 have good seniority for any interaction~\cite{Escuderos2006a}.
 They have vanishing matrix elements with the other $v=2$ and $v=4$
 states, irrespective of two-body interactions used. In other words,
those two special $v=4$ states
are uniquely specified and are eigenstates of any two-body interaction. In the following we those special states  and the $v=4$ states orthogonal to them as $|\alpha\rangle$
 and $|\beta\rangle$, respectively.
Detailed descriptions of the problem can be found in Refs. \cite{doi:10.1080/10619127.2014.883479,VanIsacker201473,Escuderos2006a,PhysRevC.75.064305,Qi2012a,PhysRevLett.100.052501,Qi2011a,Leviatan201193,doi:10.1142/S021830131101751X}.
An analytical proof for such partial conservation of seniority is
also given in Refs. \cite{Qi2011a,Qi2012a}. 

In this letter we will show that the existence of partial conservation
of seniority in $j=9/2$ shells plays an essential role in our understanding
of the electric quadrupole transitions of the nuclei involved. Another
important objective of this paper is to explore how the unique states
mentioned above, which are defined for single-$j$ systems, are influenced
by configuration mixing from other neighboring subshells. We will
show that a sharp transition from pure seniority coupling to significant
mixing between the $v=2$ and $v=4,\alpha$ states may be induced
by the cross-orbital non-diagonal interaction matrix elements. Such
strong mixture is essential to explain the observed E2 transition
properties of $N=50$ isotones $^{96}$Pd and $^{94}$Ru. In a similar
context, Ref. \cite{Escuderos2006} discussed briefly the consequences
of multi-shell calculations for states that are degenerate within
a single-$j$ shell. 

We will focus on the lightest semi-magic nuclei that involve a $j=9/2$
orbital. These include the Ni isotopes between $N=40$ and 50 and
$N=50$ isotones between $Z=40$ and 50 (see Ref. \cite{Faestermann201385}
for a review on the structure of nuclei in this region). Those nuclei
are expected to be dominated by the coupling within the $0g_{9/2}$
shell but the contribution from other neighboring orbitals (including
$1p_{1/2},\,1p_{3/2},\,0f_{5/2}$) may also play an important role.
A microscopic description of the many-body wave function is provided
by the shell model full configuration interaction approach where the
superposition of a sufficiently large number of many-body basis states
within a given valence model space are considered. As for the $N=50$
isotones, there has been many studies within the model spaces that
include the $g_{9/2}$ orbital, the $1p_{1/2}0g_{9/2}$ orbitals as
well as the $0f_{5/2}1p_{3/2}1p_{1/2}0g_{9/2}$ orbitals. All our
calculations below are done numerically within the full shell model
framework with exact diagonalization. 

\begin{figure}
\includegraphics{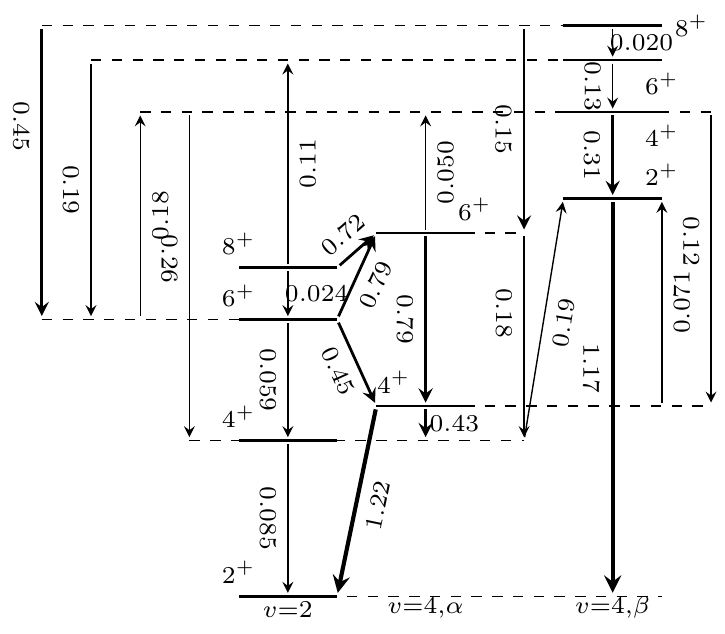}\caption{The E2 transition strengths relative to $B(E2;2_{1}^{+}\rightarrow0_{1}^{+})$
(denoted as $B_{20}$) for a system with four particles (holes) in
$j=9/2$ shell calculated using a seniority-conserving interaction.
The $0^{+}$ states and few weakest transitions are not shown for
simplicity. One has $B(E2;2_{1}^{+}\rightarrow0_{2}^{+})=0.044B_{20}$
and $B(E2;2_{2}^{+}\rightarrow0_{2}^{+})=0.19B_{20}$. The two $v=4,$
$\alpha$ states do not mix with others for any $g_{9/2}$ interaction.\label{fig:The-E2-transition}}
\end{figure}

We have done calculations for different $(g_{9/2})^{4}$ systems within
the $g_{9/2}$ orbital. The calculations are exactly the same for
the spectra and E2 transition properties of the four-particle/four
hole systems $^{94}$Ru and $^{96}$Pd (and $^{72}$Ni and $^{74}$Ni).
In Fig. \ref{fig:The-E2-transition} a detailed calculation is given
on the relative E2 transition strengths for a $(9/2)^{4}$ system
calculated with a seniority-conserving (SC) interaction. Part of the
results may also be found in Ref. \cite{PhysRevLett.100.052501}. The E2 transition matrix elements between states
with the same seniority is related to each other as $\langle j^n vI ||E2||j^n vI'\rangle=(2j+1-2n)/(2j+1-2v)\langle j^v vI ||E2||j^v vI'\rangle$.
As a result, the E2 transitions involve $v=2$ are mostly weak.
On the other hand, as indicated in Fig. \ref{fig:The-E2-transition}, the E2 transitions between the two special
$v=4,\alpha$ states and between those states are strong and are proportional to $B(E2;2^+_1\rightarrow 0^+_1)$. The transitions between those $v=4$ states and the $v=2$ states are also expected to be strong. 
However, those special states are weakly connected to the other
$v=4$ states. 

The lowest-lying spectra for such semi-magic nuclei are usually dominated
by low seniority states. The seniority coupling is also associated
with the existence of long-lived isomeric states with aligned spin
$I=2j-1$ and seniority $v=2$ in relation to the diminishing energy
gap between the isomer and the $I=2j-3$ state and the suppressed
E2 transition between the two. The suppression is expected to be maximum
when the subshell is half-occupied. A systematic study on those E2
transitions may be found, e.g., in Ref. \cite{Ressler2004}. The situation
for $(9/2)^{4}$ systems can be much more complicated since the two
$\alpha$ states are also expected to have rather low excitation energies.
Analytic expressions have been derived for their energies which depend
on the strengths of the matrix elements $\langle0g^2_{9/2}|V|0g^2_{9/2}\rangle_{J}$
with $J\neq0$ \cite{PhysRevLett.100.052501}. 

\begin{figure}
\includegraphics{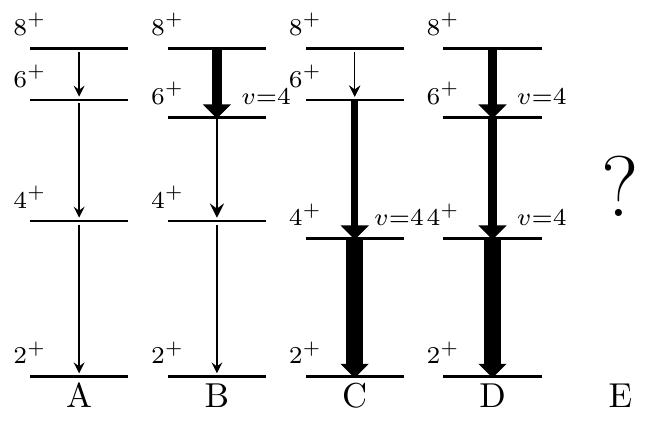}\caption{Illustration on E2 transitions for the yrast states of a $(9/2)^{4}$ system in different
scenarios based on E2 transitions from Fig. \ref{fig:The-E2-transition}: A. All lowest excited states are dominated by seniority
$v=2$ configurations with suppressed E2 transitions below them; B.
The special $v=4$, $6^{+}$ state becomes yrast with a large $B(E2;8_{1}^{+}\rightarrow6_{1}^{+})$
value, in which situation the $8_{1}^{+}$ state may not be isomeric;
C. Similar to B but with the special $v=4$, $4^{+}$ state becomes
yrast; D. Both special $v=4$, $4^{+}$ and $6^{+}$ states become
yrast where a collective-like strong inband E2 transition pattern
is formed. E. One may wonder if it is possible to have a strong mixture
between the $v=2$ and 4 states (see text for details). \label{fig:E2-transitions-for}}
\end{figure}

A schematic plot for the influence of the relative positions of low-lying
states on the yrast E2 transition properties are shown in Fig. \ref{fig:E2-transitions-for}.
The low-lying spectroscopy of $^{72}$Ni including the $4_{2}^{+}$
, $6_{2}^{+}$ and $8_{1}^{+}$ states was reported in Ref. \cite{PhysRevC.93.034328}.
The $B(E2;4_{1}^{+}\rightarrow2_{1}^{+})$ value for $^{72}$Ni was
measured to be 50(9) e$^{2}$fm$^{4}$ in Ref. \cite{PhysRevLett.116.122502},
which indicates that the $4_{1}^{+}$ state may be mostly of seniority
$v=4$ (see, also, Fig. 4 in Ref. \cite{VanIsacker201473}). As a
result, the $8_{1}^{+}$ states in $^{72,74}$Ni are not expected
to be isomeric \cite{Mazzocchi200545,PhysRevC.68.044304,PhysRevC.70.044314}. 

A tentative search for the $6_{2}^{+}$ state in $^{94}$Ru was reported
in Ref. \cite{PhysRevC.75.047302}. For $^{94}$Ru and $^{96}$Pd,
the two $\alpha$ states are expected to be just above the yrast $I=4$
and 8 states, respectively, in most of our calculations. The $4_{2}^{+}$
states in $^{94}$Ru and $^{96}$Pd were also predicted to be lower
than $6_{1}^{+}$ in the $pg$ calculations in Refs. \cite{GLOECKNER1974477,1402-4896-31-1-006}.
Restricted calculations with the interactions from Ref. \cite{Hon09,PhysRevC.70.044314}
predict the two $v=4$ states to be yrast. When extended to the full
$fpg$ space, the $6_{2}^{+}$ state is calculated to be 35 keV above
the $8_{1}^{+}$ state with the jun45 interaction \cite{Hon09}. 

The nucleus $^{94}$Ru has an $8^{+}$ isomer at 2.644 MeV with a
half-life of 71 \textgreek{m}s \cite{nudat}. The isomeric character
of this level is a consequence of the significantly suppressed E2
decay and the small energy difference with the $6^{+}$ level below
it. The E2 transition probabilities in $^{94}$Ru have been calculated
in Refs. \cite{0305-4616-6-3-010,GLOECKNER1972597,BALL1969182,SCHNEIDER1975103,GLOECKNER1974477,JAKLEVIC1969179}.

The existence of the uniquely defined $v=4$, $\alpha$ states makes
it possible to understand the suppression of $B(E2;8_{1}^{+}\rightarrow6_{1}^{+})$
in $^{94}$Ru from a very simple perspective. Since those two states
do not mix with others, one can write the wave functions of the seniority
$v'=2$ (here one uses $v'$ to denote states with mixed seniorities
but are dominated by the configuration with seniority $v$), $6^{+}$
and $8^{+}$ states as $|j^4,I\rangle_1=\beta_2^I|j^4,v=2,I\rangle+\beta_4^I|j^4,v=4,\beta,I\rangle$ where $\beta_v^I$ denotes the amplitude.
By taking $B(E2;I_{i}\rightarrow I_{f})=e_{\mathrm{eff}}^{2}|M(E2;I_{i}\rightarrow I_{f})|^{2}/(2I_{i}+1)$
and $M_{v_{1}v_{2}}=M(E2;8^{+}(v_{1})\rightarrow6^{+}(v_{2})),$ one
can calculate the transition element as
\begin{eqnarray}
M(E2;8_{1}^{+} & \rightarrow & 6_{1}^{+})\\
 & = & \beta_{2}^{8}\beta_{2}^{6}M_{22}+[\beta_{4}^{8}\beta_{2}^{6}M_{42}+\beta_{2}^{8}\beta_{4}^{6}M_{24}]+\beta_{4}^{8}\beta_{4}^{6}M_{44},\nonumber 
\end{eqnarray}
where $M_{22}$ is of positive value and the rest are negative. One
should expect the absolute values of $\beta_{2}^{I}$ to be much larger
than that of $\beta_{4}^{I}$ since the $v=4,\,\beta$ states lie
at rather high excitation energies. Moreover, as indicated in Fig.
\ref{fig:The-E2-transition}, the absolute values for $M_{22}$ and
$M_{44}$ are much smaller than the other two. As a result, the suppression
of the transition should be mostly due to the cancellation of the
first and middle two terms in the bracket where $\beta_{4}^{I}$ should
have the same sign as $\beta_{2}^{I}$. 

\begin{figure}
\includegraphics{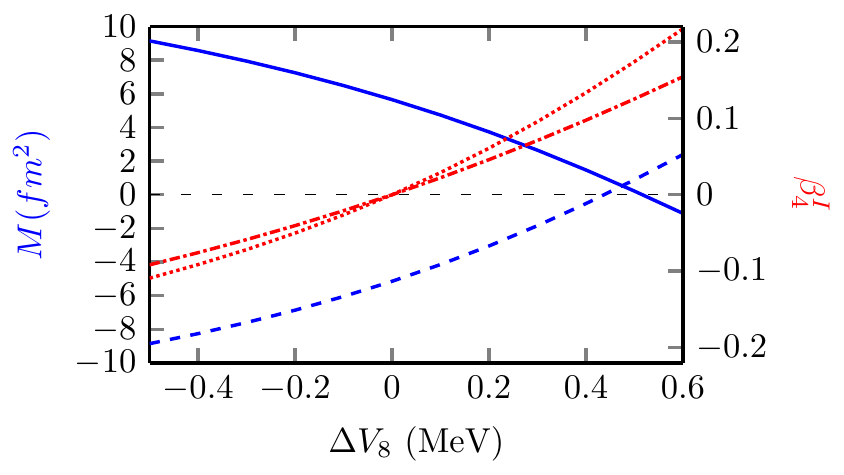}\caption{Influence of the isospin-non-conserving matrix element on the wave
functions of the $8_{1}^{+}(v'=2)$ and $6_{1}^{+}(v'=2)$ states
in $^{94}$Ru and on the tansition matrix element $M(E2;8_{1}^{+}\rightarrow6_{1}^{+})$
(blue solid line) and $M(E2;8_{2}^{+}\rightarrow6_{3}^{+})$ (blue
dashed line). Calculations are done by shifting the strength of the
matrix element $V_{8}$ of the SC interaction by an amount $\Delta v$.
The red solid and red dashed lines correspond to $\beta_{4}^{6}$
and $\beta_{4}^{8}$ values for the $v'=2$ states where it is assumed
$\beta_{2}^{I}>0.$ Those two amplitudes change sign at $\Delta v=0$.
\label{fig:A-schematic-plot}}
\end{figure}

To illustrate the influence of the seniority mixing on the E2 transition
property, in Fig. \ref{fig:A-schematic-plot} I calculated the wave
functions and transition matrix element by varying the seniority-non-conserving
interaction matrix element $V_{SNC}=65V_{2}-315V_{4}+403V_{6}-153V_{8}$.
Only $M_{22}$ contributes for $V_{SNC}$ (or $\Delta V_{8}$)$=0$.
$\beta_{4}^{I}$ show finite values with the same sign as $\beta_{2}^{I}$
for negative $V_{SNC}$, which eventually lead to a full cancellation
of $M(E2)$. 

\begin{figure*}
\includegraphics[scale=1.2]{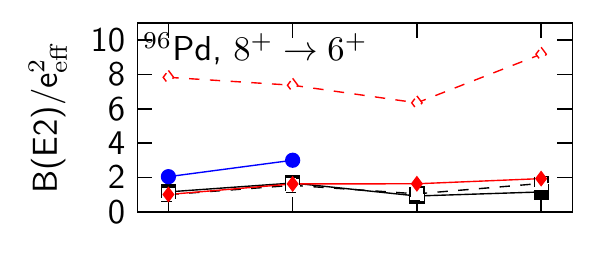}\includegraphics[scale=1.2]{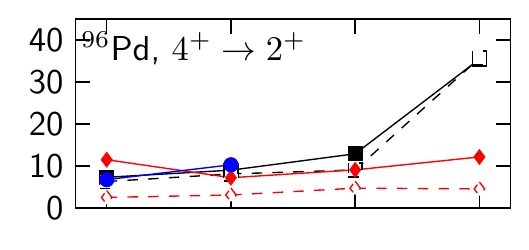}

\includegraphics[scale=1.2]{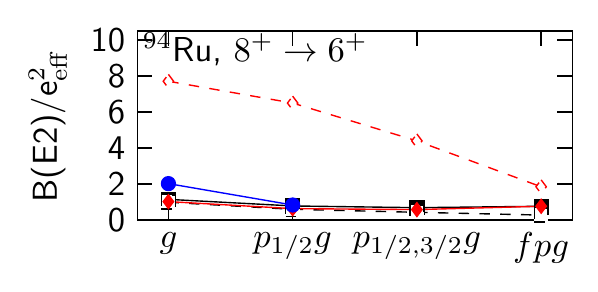}\includegraphics[scale=1.2]{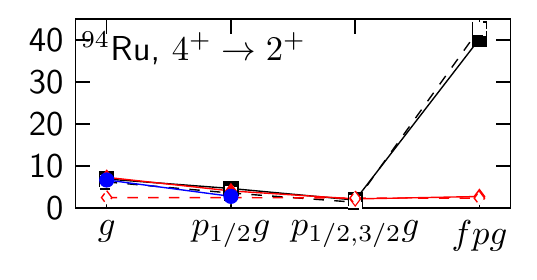}\caption{$8_{v'=2}^{+}\rightarrow6_{v'=2}^{+}$ and $4_{v'=2}^{+}\rightarrow2_{v'=2}^{+}$
E2 transition strengths (divided by the square of the effective charge,
in fm$^{4})$ for $^{94}$Ru, $^{96}$Pd calculated in different model
spaces with effective interactions from Refs. \cite{Hon09} (square),
\cite{PhysRevC.70.044314} (diamond) and \cite{1402-4896-31-1-006}
(circle). The open symbols connected by dashed lines correspond to
the calculations for $^{72,74}$Ni. The experimental $B(E2)$ values
are 0.090 (5) and 8.9 (13) $\textrm{\ensuremath{e^{2}}}\textrm{f\ensuremath{\textrm{m}^{4}}}$,
respectively, for $^{94}$Ru, $^{96}$Pd \cite{doi:10.1142/9789812702401_0042,nudat}.
\label{fig:5}}
\end{figure*}

As indicated in Fig. \ref{fig:A-schematic-plot}, the transition $8_{2}^{+}(v'=4)\rightarrow6_{3}^{+}(v'=4)$
will also be suppressed for the same reason. On the other hand, the
$g_{9/2}$ matrix elements from the effective interactions for Ni
isotopes (jj44Ni and jj44b)\cite{PhysRevC.70.044314} also show rather
large seniority non-conserving matrix element but with a different
sign. In that case, as shown in Fig. \ref{fig:5}, the predicted $B(E2;8_{v'=2}^{+}\rightarrow6_{v'=2}^{+})$
values for $^{72,74}$Ni are much larger than those from other interactions
and no cancellation is expected. 

In order to explore the influence of the neighboring orbitals, in Fig.
\ref{fig:5} I have done calculations with different effective interactions
on the transitions $8_{v'=2}^{+}\rightarrow6_{v'=2}^{+}$ and $4_{v'=2}^{+}\rightarrow2_{v'=2}^{+}$
by gradually enlarging the model space. No significant influence from
the mixture with those orbitals is seen for the $8_{v'=2}^{+}\rightarrow6_{v'=2}^{+}$
transition in $^{96}$Pd (and $^{74}$Ni). Moreover, the opening of
the $N/Z=50$ shell closures is not expected to influence the E2 transitions
in the $N=50$ isotones in a significant manner. On the other hand,
if the model space is extended to include the $p_{1/2}$ orbital,
the transitions for $^{94}$Ru and $^{72}$Ni can be influenced by
the mixture between $|g_{9/2}^{-6}\rangle J$ and $|p_{1/2}^{-2}g_{9/2}^{-4}\rangle_{J}$
configurations. This is related to the cancellation as induced by
the four-particle and four-hole natures of the two configurations.
Such kind of cancellation does not happen for $^{96}$Pd and $^{74}$Ni.
This is partly responsible for the fact that the observed transition
probability $B(E2;8_{1}^{+}\rightarrow6_{1}^{+})$ for $^{96}$Pd
is nearly 100 times larger than that of $^{94}$Ru. It is also noticed
that, for the same reason, the measured $B(E2;6_{1}^{+}\rightarrow4_{1}^{+})$
value for $^{96}$Pd \cite{doi:10.1142/9789812702401_0042} is more
than eight times larger than that of $^{94}$Ru.

In Ref. \cite{doi:10.1142/9789812702401_0042}, the $B(E2;4_{1}^{+}\rightarrow2_{1}^{+})$
value for $^{96}$Pd was measured to be as small as 3.8 $\textrm{\textrm{\ensuremath{\textrm{e}^{2}}}}\textrm{f\ensuremath{\textrm{m}^{4}}}$,
which is significantly suppressed by roughly a factor of seven in
comparison with that predicted by a SC interaction. In contrast to
those for $\rightarrow6_{1}^{+}$ and $6_{1}^{+}\rightarrow4_{1}^{+}$,
that value is expected to be significantly smaller than the that for
$^{94}$Ru where the lower limit for $B(E2;4_{1}^{+}\rightarrow2_{1}^{+})$
is suggested to be as large as 46 $\textrm{\ensuremath{\textrm{e}^{2}}f\ensuremath{\textrm{m}^{4}}}$.
Such an anomalous suppression can not be reproduced by calculations
within the single $g_{9/2}$ shell but should be related to the mixing
with other shells. In the following I will show that such anomalous
transition is related to the unexpected mixture between $v=2$ and
$v=4,\alpha$ which is induced by cross-orbital non-diagonal matrix
elements of the two-body interaction. A detailed analysis on all related
transitions will be presented in a forthcoming paper. Moreover, a
dramatic increase in the $B(E2;4_{v'=2}^{+}\rightarrow2_{v'=2}^{+}$)
values of $^{96}$Pd and $^{94}$Ru is seen in Fig. \ref{fig:5} for
calculations with the jun45 interaction when the model space is extended
to include $f_{5/2}$. Our detailed analysis of the corresponding
wave functions shows that this calculated abrupt change is also related
to the configuration mixing within $g_{9/2}$ induced by non-diagonal
matrix elements involving $f_{5/2}$. 

\begin{figure}
\includegraphics{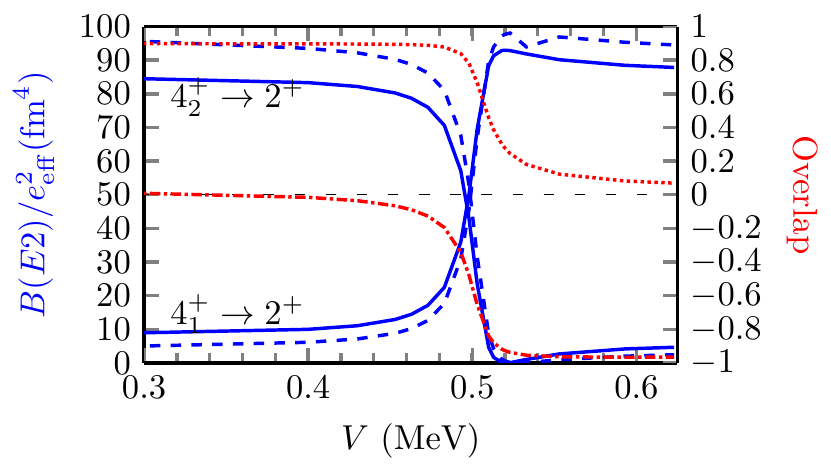}\caption{E2 transition strengths (solid lines) for the transitions $4_{1,2}^{+}\rightarrow2_{1}^{+}$
in $^{96}$Pd calculated in a minimal model space $p_{1/2,3/2}g$
calculated by varying the strength of the non-diagonal matrix element
$V_{p_{3/2}p_{3/2}g_{9/2}g_{9/2}}^{J=2}$. The dashed lines correspond
to the transition from $4_{1,2}^{+}$ to the state $|g_{9/2}^{-4},v=2,I=2\rangle$.
The dotted and dash-dotted lines (red) show the overlaps between $4_{1}^{+}$
and the seniority $v=2$ and $v=4,\alpha$ states. Calculations are
done with the jun45 effective Hamiltonian by allowing at most two
particles/holes in $p_{3/2}$. The original value of the matrix element
is 0.453 MeV  while a sharp transition occurs between 0.46 and 0.52
MeV where the main component of $4_{1}^{+}$ ($4_{2}^{+}$) change
from seniority 2 (4) to 4 (2). The transition $4_{2}^{+}(v'=2)\rightarrow2_{1}^{+}$
vanish with $V_{p_{3/2}p_{3/2}g_{9/2}g_{9/2}}^{J=2}\approx0.52$ MeV.
With this interaction strength, a strong mixture between the $v=2$
and $\alpha$ configurations is still expected for $4_{1,2}^{+}$
in $^{94}$Ru.}
\label{4to2}
\end{figure}

The overlaps between the two special $I=4$ and 6, $\alpha$ states
with the states constructed from the coupling of two $J=2$ pairs
$|j_{J=2}^{2}\otimes j_{J=2}^{2}\rangle_{I=4}$ and two $J=2$ and
$J=4$ pairs $|j_{J=2}^{2}\otimes j_{J=4}^{2}\rangle_{I=6}$ are as
large as $\alpha^I=10\sqrt{255}/\sqrt{25591}\approx0.9982$ and $2\sqrt{6783}/\sqrt{27257}\approx0.9977$,
respectively. It means that the cross-orbital configurations of the
form $|(j_{1}j_{2})\otimes(g_{9/2})^{2}\rangle_{I=4,6}$ may overlap
largely with the $v=4,\alpha$ states through the non-diagonal matrix
elements $V_{j_{1}j_{2}g_{9/2}g_{9/2}}^{J=2,4}$. Those configurations
also show non-zero non-diagonal matrix elements with the $v=2$ states.
These matrix elements lead to a co-existence of the two $v=2$ and
4 configurations which does not happen in calculations within the
$g_{9/2}$. 

As for $I=4,$ it is found that the non-diagonal matrix elements with
$j_{1}j_{2}=p_{3/2}^{2}$, $p_{1/2}p_{3/2}$, $p_{1/2}f_{5/2}$, $p_{3/2}f_{5/2}$
coupled to $J=2$ can indeed induce significant mixture between the
$v=2$ and $v=4,\alpha$ states. But it happens only in a relatively
small window of strengths for the two-body matrix elements. As for
calculations in Fig. \ref{fig:5}, only those from the jun45 interaction
(more exactly, the $V_{p_{3/2}f_{5/2}g_{9/2}g_{9/2}}^{J=2}$ element)
fall in that window. That is why there is no abrupt change seen in
other calculations. It should also be mentioned that those non-diagonal
matrix elements $V_{j_{1}j_{2}g_{9/2}g_{9/2}}^{J=2,4}$ have very
limited influence on the energies of the states of concern. In relation
to that, it has always been a challenging task to pin down the sign
and the strengths of the non-diagonal interaction matrix elements
for the shell-model Hamiltonian which may be approximated from realistic
nucleon-nucleon potentials.

In Fig. \ref{4to2} I evaluated the overlaps between the calculated
wave functions and the $v=2$ and $v=4,\alpha$ for the first two
$4^{+}$ states in $^{96}$Pd in a model space containing orbitals
$p_{1/2,}p_{3/2}$ and $g_{9/2}$. That is the minimal space that
can induce significant mixture between the two $v=2$ and 4 configurations.
As indicated in Figs. \ref{fig:5} and \ref{4to2}, no significant
mixture between the two components is seen in the calculation with
the original jun45 interaction since the $V_{p_{3/2}p_{3/2}g_{9/2}g_{9/2}}^{J=2}$
interaction is slightly outside the strength window. But a strong
mixture between the two $v=2$ and 4 configurations is expected for
both $4_{1,2}^{+}$ if the interaction got more repulsive.

The transition pattern shown in Fig. \ref{4to2} gives us an unique
opportunity to understand the $4^{+}\rightarrow2^{+}$ E2 transitions
of $^{96}$Pd and $^{94}$Ru as measured in Ref. \cite{doi:10.1142/9789812702401_0042,mach2017}:
The E2 transition in $^{96}$Pd corresponds to a vanishing $4_{v'=2}^{+}\rightarrow2_{v'=2}^{+}$
transition seen in right-hand side of Fig. \ref{4to2} while the large
E2 transition in $^{94}$Ru indicates that the nucleus is indeed located
in the transitional region where the transition strength is very sensitive
to the mixture of the two configurations. 

To summarize, in this work I present a novel analysis on the electric
quadrupole transition properties of semi-magic nuclei with four particles
or four holes in the $g_{9/2}$ orbital from a partial seniority conservation
perspective. This is related to the existence of uniquely defined
$v=4$ states which, for systems within a $j=9/2$ subshell, do not
mix with other states. It is shown that the diminishing $B(E2;8_{1}^{+}\rightarrow6_{1}^{+})$
in $^{94}$Ru can be mostly understood as the cancellation between
few terms induced by the seniority-non-conserving interaction. Moroever,
I studied the influence of the neighboring $1p_{1/2},\,1p_{3/2},\,0f_{5/2}$
orbitals. It is seen that the cross-orbital interaction matrix elements
can induce significant mixture between the $v=2$ and the unique
$\alpha$states. The limited experimental information available do
indicate that such a sharp phase transition can be seen in nuclei
like $^{96}$Pd and $^{94}$Ru. In the future, besides the measurement
on the predicted states and E2 transitions mentioned in the present
work, it can also be of great interest to explore other $j=9/2$  nuclei,
including the $N=82$ isotones and neutron-rich Pb isotopes, with
different two-body interaction strengths and different neighboring
orbitals to get a better understanding of such phase transitions.
\begin{acknowledgments}
This work is supported by the Swedish Research Council (VR) under
grant Nos. 621-2012-3805, and 621-2013-4323 and the G\"oran Gustafsson
foundation. I also thank B. Cederwall for discussions. Computational
support provided by the Swedish National Infrastructure for Computing
(SNIC) at PDC, KTH, Stockholm is also acknowledged.
\end{acknowledgments}

%\bibliographystyle{apsrev4-1}
%\bibliography{/Users/chong/mypapers/mybib}

\begin{thebibliography}{36}%
\makeatletter
\providecommand \@ifxundefined [1]{%
 \@ifx{#1\undefined}
}%
\providecommand \@ifnum [1]{%
 \ifnum #1\expandafter \@firstoftwo
 \else \expandafter \@secondoftwo
 \fi
}%
\providecommand \@ifx [1]{%
 \ifx #1\expandafter \@firstoftwo
 \else \expandafter \@secondoftwo
 \fi
}%
\providecommand \natexlab [1]{#1}%
\providecommand \enquote  [1]{``#1''}%
\providecommand \bibnamefont  [1]{#1}%
\providecommand \bibfnamefont [1]{#1}%
\providecommand \citenamefont [1]{#1}%
\providecommand \href@noop [0]{\@secondoftwo}%
\providecommand \href [0]{\begingroup \@sanitize@url \@href}%
\providecommand \@href[1]{\@@startlink{#1}\@@href}%
\providecommand \@@href[1]{\endgroup#1\@@endlink}%
\providecommand \@sanitize@url [0]{\catcode `\\12\catcode `\$12\catcode
  `\&12\catcode `\#12\catcode `\^12\catcode `\_12\catcode `\%12\relax}%
\providecommand \@@startlink[1]{}%
\providecommand \@@endlink[0]{}%
\providecommand \url  [0]{\begingroup\@sanitize@url \@url }%
\providecommand \@url [1]{\endgroup\@href {#1}{\urlprefix }}%
\providecommand \urlprefix  [0]{URL }%
\providecommand \Eprint [0]{\href }%
\providecommand \doibase [0]{http://dx.doi.org/}%
\providecommand \selectlanguage [0]{\@gobble}%
\providecommand \bibinfo  [0]{\@secondoftwo}%
\providecommand \bibfield  [0]{\@secondoftwo}%
\providecommand \translation [1]{[#1]}%
\providecommand \BibitemOpen [0]{}%
\providecommand \bibitemStop [0]{}%
\providecommand \bibitemNoStop [0]{.\EOS\space}%
\providecommand \EOS [0]{\spacefactor3000\relax}%
\providecommand \BibitemShut  [1]{\csname bibitem#1\endcsname}%
\let\auto@bib@innerbib\@empty
%</preamble>
\bibitem [{\citenamefont {Talmi}(1993)}]{tal93}%
  \BibitemOpen
  \bibfield  {author} {\bibinfo {author} {\bibfnamefont {I.}~\bibnamefont
  {Talmi}},\ }\href@noop {} {\emph {\bibinfo {title} {{Simple Models of Complex
  Nuclei: The Shell Model and Interacting Boson Model}}}}\ (\bibinfo
  {publisher} {Harwood Academic Publishers},\ \bibinfo {year}
  {1993})\BibitemShut {NoStop}%
\bibitem [{\citenamefont {Morales}\ \emph {et~al.}(2016)\citenamefont {Morales}
  \emph {et~al.}}]{PhysRevC.93.034328}%
  \BibitemOpen
  \bibfield  {author} {\bibinfo {author} {\bibfnamefont {A.~I.}\ \bibnamefont
  {Morales}} \emph {et~al.},\ }\href {\doibase 10.1103/PhysRevC.93.034328}
  {\bibfield  {journal} {\bibinfo  {journal} {Phys. Rev. C}\ }\textbf {\bibinfo
  {volume} {93}},\ \bibinfo {pages} {034328} (\bibinfo {year}
  {2016})}\BibitemShut {NoStop}%
\bibitem [{\citenamefont {Watanabe}\ \emph {et~al.}(2013)\citenamefont
  {Watanabe} \emph {et~al.}}]{PhysRevLett.111.152501}%
  \BibitemOpen
  \bibfield  {author} {\bibinfo {author} {\bibfnamefont {H.}~\bibnamefont
  {Watanabe}} \emph {et~al.},\ }\href@noop {} {\bibfield  {journal} {\bibinfo
  {journal} {Phys. Rev. Lett.}\ }\textbf {\bibinfo {volume} {111}},\ \bibinfo
  {pages} {152501} (\bibinfo {year} {2013})}\BibitemShut {NoStop}%
\bibitem [{\citenamefont {Simpson}\ \emph {et~al.}(2014)\citenamefont {Simpson}
  \emph {et~al.}}]{PhysRevLett.113.132502}%
  \BibitemOpen
  \bibfield  {author} {\bibinfo {author} {\bibfnamefont {G.~S.}\ \bibnamefont
  {Simpson}} \emph {et~al.},\ }\href {\doibase 10.1103/PhysRevLett.113.132502}
  {\bibfield  {journal} {\bibinfo  {journal} {Phys. Rev. Lett.}\ }\textbf
  {\bibinfo {volume} {113}},\ \bibinfo {pages} {132502} (\bibinfo {year}
  {2014})}\BibitemShut {NoStop}%
\bibitem [{\citenamefont {Gottardo}\ \emph {et~al.}(2012)\citenamefont
  {Gottardo} \emph {et~al.}}]{PhysRevLett.109.162502}%
  \BibitemOpen
  \bibfield  {author} {\bibinfo {author} {\bibfnamefont {A.}~\bibnamefont
  {Gottardo}} \emph {et~al.},\ }\href {\doibase 10.1103/PhysRevLett.109.162502}
  {\bibfield  {journal} {\bibinfo  {journal} {Phys. Rev. Lett.}\ }\textbf
  {\bibinfo {volume} {109}},\ \bibinfo {pages} {162502} (\bibinfo {year}
  {2012})}\BibitemShut {NoStop}%
  \bibitem [{\citenamefont {Rowe}\ and\ \citenamefont
  {Rosensteel}(2001)}]{PhysRevLett.87.172501}%
  \BibitemOpen
  \bibfield  {author} {\bibinfo {author} {\bibfnamefont {D.~J.}\ \bibnamefont
  {Rowe}}\ and\ \bibinfo {author} {\bibfnamefont {G.}~\bibnamefont
  {Rosensteel}},\ }\href@noop {} {\bibfield  {journal} {\bibinfo  {journal}
  {Phys. Rev. Lett.}\ }\textbf {\bibinfo {volume} {87}},\ \bibinfo {pages}
  {172501} (\bibinfo {year} {2001})}\BibitemShut {NoStop}%
\bibitem [{\citenamefont {Rosensteel}\ and\ \citenamefont
  {Rowe}(2003)}]{PhysRevC.67.014303}%
  \BibitemOpen
  \bibfield  {author} {\bibinfo {author} {\bibfnamefont {G.}~\bibnamefont
  {Rosensteel}}\ and\ \bibinfo {author} {\bibfnamefont {D.~J.}\ \bibnamefont
  {Rowe}},\ }\href@noop {} {\bibfield  {journal} {\bibinfo  {journal} {Phys.
  Rev. C}\ }\textbf {\bibinfo {volume} {67}},\ \bibinfo {pages} {014303}
  (\bibinfo {year} {2003})}\BibitemShut {NoStop}%
\bibitem [{\citenamefont {Qi}(2010)}]{Qi10}%
  \BibitemOpen
  \bibfield  {author} {\bibinfo {author} {\bibfnamefont {C.}~\bibnamefont
  {Qi}},\ }\href@noop {} {\bibfield  {journal} {\bibinfo  {journal} {Phys. Rev.
  C}\ }\textbf {\bibinfo {volume} {81}},\ \bibinfo {pages} {034318} (\bibinfo
  {year} {2010})}\BibitemShut {NoStop}%
\bibitem [{\citenamefont {Qi}\ \emph {et~al.}(2010)\citenamefont {Qi},
  \citenamefont {Wang}, \citenamefont {Xu}, \citenamefont {Liotta},
  \citenamefont {Wyss},\ and\ \citenamefont {Xu}}]{PhysRevC.82.014304}%
  \BibitemOpen
  \bibfield  {author} {\bibinfo {author} {\bibfnamefont {C.}~\bibnamefont
  {Qi}}, \bibinfo {author} {\bibfnamefont {X.~B.}\ \bibnamefont {Wang}},
  \bibinfo {author} {\bibfnamefont {Z.~X.}\ \bibnamefont {Xu}}, \bibinfo
  {author} {\bibfnamefont {R.~J.}\ \bibnamefont {Liotta}}, \bibinfo {author}
  {\bibfnamefont {R.}~\bibnamefont {Wyss}}, \ and\ \bibinfo {author}
  {\bibfnamefont {F.~R.}\ \bibnamefont {Xu}},\ }\href@noop {} {\bibfield
  {journal} {\bibinfo  {journal} {Phys. Rev. C}\ }\textbf {\bibinfo {volume}
  {82}},\ \bibinfo {pages} {014304} (\bibinfo {year} {2010})}\BibitemShut
  {NoStop}%
  \bibitem [{\citenamefont {Van~Isacker}\ and\ \citenamefont
  {Heinze}(2014)}]{VanIsacker201473}%
  \BibitemOpen
  \bibfield  {author} {\bibinfo {author} {\bibfnamefont {P.}~\bibnamefont
  {Van~Isacker}}\ and\ \bibinfo {author} {\bibfnamefont {S.}~\bibnamefont
  {Heinze}},\ }\href@noop {} {\bibfield  {journal} {\bibinfo  {journal} {Annals
  of Physics}\ }\textbf {\bibinfo {volume} {349}},\ \bibinfo {pages} {73 }
  (\bibinfo {year} {2014})}\BibitemShut {NoStop}%
  \bibitem{Gross76}R.Gross, A.Frenkel, Nucl.Phys. A267, 85 (1976).
\bibitem [{\citenamefont {Escuderos}\ and\ \citenamefont
  {Zamick}(2006)}]{Escuderos2006a}%
  \BibitemOpen
  \bibfield  {author} {\bibinfo {author} {\bibfnamefont {A.}~\bibnamefont
  {Escuderos}}\ and\ \bibinfo {author} {\bibfnamefont {L.}~\bibnamefont
  {Zamick}},\ }\href@noop {} {\bibfield  {journal} {\bibinfo  {journal} {Phys.
  Rev. C}\ }\textbf {\bibinfo {volume} {73}}, 044302 (\bibinfo {year}
  {2006})}\BibitemShut {NoStop}%
\bibitem [{\citenamefont {Zamick}(2007)}]{PhysRevC.75.064305}%
  \BibitemOpen
  \bibfield  {author} {\bibinfo {author} {\bibfnamefont {L.}~\bibnamefont
  {Zamick}},\ }\href@noop {} {\bibfield  {journal} {\bibinfo  {journal} {Phys.
  Rev. C}\ }\textbf {\bibinfo {volume} {75}},\ \bibinfo {pages} {064305}
  (\bibinfo {year} {2007})}\BibitemShut {NoStop}%
\bibitem [{\citenamefont {Van~Isacker}\ and\ \citenamefont
  {Heinze}(2008)}]{PhysRevLett.100.052501}%
  \BibitemOpen
  \bibfield  {author} {\bibinfo {author} {\bibfnamefont {P.}~\bibnamefont
  {Van~Isacker}}\ and\ \bibinfo {author} {\bibfnamefont {S.}~\bibnamefont
  {Heinze}},\ }\href@noop {} {\bibfield  {journal} {\bibinfo  {journal} {Phys.
  Rev. Lett.}\ }\textbf {\bibinfo {volume} {100}},\ \bibinfo {pages} {052501}
  (\bibinfo {year} {2008})}\BibitemShut {NoStop}%
\bibitem [{\citenamefont {Qi}(2011)}]{Qi2011a}%
  \BibitemOpen
  \bibfield  {author} {\bibinfo {author} {\bibfnamefont {C.}~\bibnamefont
  {Qi}},\ }\href@noop {} {\bibfield  {journal} {\bibinfo  {journal} {Phys. Rev.
  C}\ }\textbf {\bibinfo {volume} {83}},\ \bibinfo {pages} {014307} (\bibinfo
  {year} {2011})}\BibitemShut {NoStop}%
\bibitem [{\citenamefont {Leviatan}(2011)}]{Leviatan201193}%
  \BibitemOpen
  \bibfield  {author} {\bibinfo {author} {\bibfnamefont {A.}~\bibnamefont
  {Leviatan}},\ }\href@noop {} {\bibfield  {journal} {\bibinfo  {journal}
  {Prog. Part. Nucl. Phys.}\ }\textbf {\bibinfo {volume} {66}},\ \bibinfo
  {pages} {93 } (\bibinfo {year} {2011})}\BibitemShut {NoStop}%
\bibitem [{\citenamefont {Van~Isacker}(2011)}]{doi:10.1142/S021830131101751X}%
  \BibitemOpen
  \bibfield  {author} {\bibinfo {author} {\bibfnamefont {P.}~\bibnamefont
  {Van~Isacker}},\ }\href@noop {} {\bibfield  {journal} {\bibinfo  {journal}
  {Int. J. Mod. Phys. E}\ }\textbf {\bibinfo {volume} {20}},\ \bibinfo {pages}
  {191} (\bibinfo {year} {2011})}\BibitemShut {NoStop}%
 \bibitem [{\citenamefont {Qi}\ \emph {et~al.}(2012)\citenamefont {Qi},
  \citenamefont {Xu},\ and\ \citenamefont {Liotta}}]{Qi2012a}%
  \BibitemOpen
  \bibfield  {author} {\bibinfo {author} {\bibfnamefont {C.}~\bibnamefont
  {Qi}}, \bibinfo {author} {\bibfnamefont {Z.~X.}\ \bibnamefont {Xu}}, \ and\
  \bibinfo {author} {\bibfnamefont {R.~J.}\ \bibnamefont {Liotta}},\
  }\href@noop {} {\bibfield  {journal} {\bibinfo  {journal} {Nucl. Phys. A}\
  }\textbf {\bibinfo {volume} {884}},\ \bibinfo {pages} {21} (\bibinfo {year}
  {2012})}\BibitemShut {NoStop}%
  \bibitem [{\citenamefont
  {Van~Isacker}(2014)}]{doi:10.1080/10619127.2014.883479}%
  \BibitemOpen
  \bibfield  {author} {\bibinfo {author} {\bibfnamefont {P.}~\bibnamefont
  {Van~Isacker}},\ }\href@noop {} {\bibfield  {journal} {\bibinfo  {journal}
  {Nuclear Physics News}\ }\textbf {\bibinfo {volume} {24}},\ \bibinfo {pages}
  {23} (\bibinfo {year} {2014})}\BibitemShut {NoStop}%
\bibitem [{\citenamefont {Escuderos}\ \emph {et~al.}(2006)\citenamefont
  {Escuderos}, \citenamefont {Robinson},\ and\ \citenamefont
  {Zamick}}]{Escuderos2006}%
  \BibitemOpen
  \bibfield  {author} {\bibinfo {author} {\bibfnamefont {A.}~\bibnamefont
  {Escuderos}}, \bibinfo {author} {\bibfnamefont {S.~J.~Q.}\ \bibnamefont
  {Robinson}}, \ and\ \bibinfo {author} {\bibfnamefont {L.}~\bibnamefont
  {Zamick}},\ }\href@noop {} {\bibfield  {journal} {\bibinfo  {journal} {Phys.
  Rev. C}\ }\textbf {\bibinfo {volume} {73}},\ \bibinfo {pages} {027301}
  (\bibinfo {year} {2006})}\BibitemShut {NoStop}%
\bibitem [{\citenamefont {Faestermann}\ \emph {et~al.}(2013)\citenamefont
  {Faestermann}, \citenamefont {Gorska},\ and\ \citenamefont
  {Grawe}}]{Faestermann201385}%
  \BibitemOpen
  \bibfield  {author} {\bibinfo {author} {\bibfnamefont {T.}~\bibnamefont
  {Faestermann}}, \bibinfo {author} {\bibfnamefont {M.}~\bibnamefont {Gorska}},
  \ and\ \bibinfo {author} {\bibfnamefont {H.}~\bibnamefont {Grawe}},\
  }\href@noop {} {\bibfield  {journal} {\bibinfo  {journal} {Prog. Part. Nucl.
  Phys.}\ }\textbf {\bibinfo {volume} {69}},\ \bibinfo {pages} {85 } (\bibinfo
  {year} {2013})}\BibitemShut {NoStop}%
\bibitem [{\citenamefont {Ressler}\ \emph {et~al.}(2004)\citenamefont {Ressler}
  \emph {et~al.}}]{Ressler2004}%
  \BibitemOpen
  \bibfield  {author} {\bibinfo {author} {\bibfnamefont {J.~J.}\ \bibnamefont
  {Ressler}} \emph {et~al.},\ }\href@noop {} {\bibfield  {journal} {\bibinfo
  {journal} {Phys. Rev. C}\ }\textbf {\bibinfo {volume} {69}},\ \bibinfo
  {pages} {034317} (\bibinfo {year} {2004})}\BibitemShut {NoStop}%
\bibitem [{\citenamefont {Kolos}\ \emph {et~al.}(2016)\citenamefont {Kolos}
  \emph {et~al.}}]{PhysRevLett.116.122502}%
  \BibitemOpen
  \bibfield  {author} {\bibinfo {author} {\bibfnamefont {K.}~\bibnamefont
  {Kolos}} \emph {et~al.},\ }\href {\doibase 10.1103/PhysRevLett.116.122502}
  {\bibfield  {journal} {\bibinfo  {journal} {Phys. Rev. Lett.}\ }\textbf
  {\bibinfo {volume} {116}},\ \bibinfo {pages} {122502} (\bibinfo {year}
  {2016})}\BibitemShut {NoStop}%
\bibitem [{\citenamefont {Sawicka}\ \emph {et~al.}(2003)\citenamefont {Sawicka}
  \emph {et~al.}}]{PhysRevC.68.044304}%
  \BibitemOpen
  \bibfield  {author} {\bibinfo {author} {\bibfnamefont {M.}~\bibnamefont
  {Sawicka}} \emph {et~al.},\ }\href {\doibase 10.1103/PhysRevC.68.044304}
  {\bibfield  {journal} {\bibinfo  {journal} {Phys. Rev. C}\ }\textbf {\bibinfo
  {volume} {68}},\ \bibinfo {pages} {044304} (\bibinfo {year}
  {2003})}\BibitemShut {NoStop}%
\bibitem [{\citenamefont {Lisetskiy}\ \emph {et~al.}(2004)\citenamefont
  {Lisetskiy}, \citenamefont {Brown}, \citenamefont {Horoi},\ and\
  \citenamefont {Grawe}}]{PhysRevC.70.044314}%
  \BibitemOpen
  \bibfield  {author} {\bibinfo {author} {\bibfnamefont {A.~F.}\ \bibnamefont
  {Lisetskiy}}, \bibinfo {author} {\bibfnamefont {B.~A.}\ \bibnamefont
  {Brown}}, \bibinfo {author} {\bibfnamefont {M.}~\bibnamefont {Horoi}}, \ and\
  \bibinfo {author} {\bibfnamefont {H.}~\bibnamefont {Grawe}},\ }\href@noop {}
  {\bibfield  {journal} {\bibinfo  {journal} {Phys. Rev. C}\ }\textbf {\bibinfo
  {volume} {70}},\ \bibinfo {pages} {044314} (\bibinfo {year}
  {2004})}\BibitemShut {NoStop}%
  \bibitem [{\citenamefont {Mazzocchi}\ \emph {et~al.}(2005)\citenamefont
  {Mazzocchi} \emph {et~al.}}]{Mazzocchi200545}%
  \BibitemOpen
  \bibfield  {author} {\bibinfo {author} {\bibfnamefont {C.}~\bibnamefont
  {Mazzocchi}} \emph {et~al.},\ }\href {\doibase
  http://dx.doi.org/10.1016/j.physletb.2005.07.006} {\bibfield  {journal}
  {\bibinfo  {journal} {Phys. Lett. B}\ }\textbf {\bibinfo {volume} {622}},\
  \bibinfo {pages} {45 } (\bibinfo {year} {2005})}\BibitemShut {NoStop}%
\bibitem [{\citenamefont {Mills}\ \emph {et~al.}(2007)\citenamefont {Mills},
  \citenamefont {Ressler}, \citenamefont {Austin}, \citenamefont
  {Chakrawarthy}, \citenamefont {Cross}, \citenamefont {Heinz}, \citenamefont
  {McCutchan},\ and\ \citenamefont {Strange}}]{PhysRevC.75.047302}%
  \BibitemOpen
  \bibfield  {author} {\bibinfo {author} {\bibfnamefont {W.~J.}\ \bibnamefont
  {Mills}}, \bibinfo {author} {\bibfnamefont {J.~J.}\ \bibnamefont {Ressler}},
  \bibinfo {author} {\bibfnamefont {R.~A.~E.}\ \bibnamefont {Austin}}, \bibinfo
  {author} {\bibfnamefont {R.~S.}\ \bibnamefont {Chakrawarthy}}, \bibinfo
  {author} {\bibfnamefont {D.~S.}\ \bibnamefont {Cross}}, \bibinfo {author}
  {\bibfnamefont {A.}~\bibnamefont {Heinz}}, \bibinfo {author} {\bibfnamefont
  {E.~A.}\ \bibnamefont {McCutchan}}, \ and\ \bibinfo {author} {\bibfnamefont
  {M.~D.}\ \bibnamefont {Strange}},\ }\href@noop {} {\bibfield  {journal}
  {\bibinfo  {journal} {Phys. Rev. C}\ }\textbf {\bibinfo {volume} {75}},\
  \bibinfo {pages} {047302} (\bibinfo {year} {2007})}\BibitemShut {NoStop}%
\bibitem [{\citenamefont {Gloeckner}\ and\ \citenamefont
  {Serduke}(1974)}]{GLOECKNER1974477}%
  \BibitemOpen
  \bibfield  {author} {\bibinfo {author} {\bibfnamefont {D.}~\bibnamefont
  {Gloeckner}}\ and\ \bibinfo {author} {\bibfnamefont {F.}~\bibnamefont
  {Serduke}},\ }\href@noop {} {\bibfield  {journal} {\bibinfo  {journal} {Nucl.
  Phys. A}\ }\textbf {\bibinfo {volume} {220}},\ \bibinfo {pages} {477 }
  (\bibinfo {year} {1974})}\BibitemShut {NoStop}%
\bibitem [{\citenamefont {Blomqvist}\ and\ \citenamefont
  {Rydstr\"om}(1985)}]{1402-4896-31-1-006}%
  \BibitemOpen
  \bibfield  {author} {\bibinfo {author} {\bibfnamefont {J.}~\bibnamefont
  {Blomqvist}}\ and\ \bibinfo {author} {\bibfnamefont {L.}~\bibnamefont
  {Rydstr\"om}},\ }\href@noop {} {\bibfield  {journal} {\bibinfo  {journal}
  {Physica Scripta}\ }\textbf {\bibinfo {volume} {31}},\ \bibinfo {pages} {31}
  (\bibinfo {year} {1985})}\BibitemShut {NoStop}%
\bibitem [{\citenamefont {Honma}\ \emph {et~al.}(2009)\citenamefont {Honma},
  \citenamefont {Otsuka}, \citenamefont {Mizusaki},\ and\ \citenamefont
  {Hjorth-Jensen}}]{Hon09}%
  \BibitemOpen
  \bibfield  {author} {\bibinfo {author} {\bibfnamefont {M.}~\bibnamefont
  {Honma}}, \bibinfo {author} {\bibfnamefont {T.}~\bibnamefont {Otsuka}},
  \bibinfo {author} {\bibfnamefont {T.}~\bibnamefont {Mizusaki}}, \ and\
  \bibinfo {author} {\bibfnamefont {M.}~\bibnamefont {Hjorth-Jensen}},\
  }\href@noop {} {\bibfield  {journal} {\bibinfo  {journal} {Phys. Rev. C}\
  }\textbf {\bibinfo {volume} {80}},\ \bibinfo {pages} {064323} (\bibinfo
  {year} {2009})}\BibitemShut {NoStop}%
\bibitem [{nud()}]{nudat}%
  \BibitemOpen
  \href {http://www.nndc.bnl.gov/nudat2/} {\enquote {\bibinfo {title}
  {Nudat2.6,http://www.nndc.bnl.gov/nudat2/},}\ }\BibitemShut {NoStop}%
\bibitem [{\citenamefont {Ball}\ \emph {et~al.}(1969)\citenamefont {Ball},
  \citenamefont {McGrory}, \citenamefont {Auble},\ and\ \citenamefont
  {Bhatt}}]{BALL1969182}%
  \BibitemOpen
  \bibfield  {author} {\bibinfo {author} {\bibfnamefont {J.}~\bibnamefont
  {Ball}}, \bibinfo {author} {\bibfnamefont {J.}~\bibnamefont {McGrory}},
  \bibinfo {author} {\bibfnamefont {R.}~\bibnamefont {Auble}}, \ and\ \bibinfo
  {author} {\bibfnamefont {K.}~\bibnamefont {Bhatt}},\ }\href {\doibase
  http://dx.doi.org/10.1016/0370-2693(69)90014-8} {\bibfield  {journal}
  {\bibinfo  {journal} {Phys. Lett. B}\ }\textbf {\bibinfo {volume} {29}},\
  \bibinfo {pages} {182 } (\bibinfo {year} {1969})}\BibitemShut {NoStop}%
\bibitem [{\citenamefont {Jaklevic}\ \emph {et~al.}(1969)\citenamefont
  {Jaklevic}, \citenamefont {Lederer},\ and\ \citenamefont
  {Hollander}}]{JAKLEVIC1969179}%
  \BibitemOpen
  \bibfield  {author} {\bibinfo {author} {\bibfnamefont {J.}~\bibnamefont
  {Jaklevic}}, \bibinfo {author} {\bibfnamefont {C.}~\bibnamefont {Lederer}}, \
  and\ \bibinfo {author} {\bibfnamefont {J.}~\bibnamefont {Hollander}},\ }\href
  {\doibase http://dx.doi.org/10.1016/0370-2693(69)90013-6} {\bibfield
  {journal} {\bibinfo  {journal} {Phys. Lett. B}\ }\textbf {\bibinfo {volume}
  {29}},\ \bibinfo {pages} {179 } (\bibinfo {year} {1969})}\BibitemShut
  {NoStop}%
  \bibitem [{\citenamefont {Gloeckner}\ \emph {et~al.}(1972)\citenamefont
  {Gloeckner}, \citenamefont {MacFarlane}, \citenamefont {Lawson},\ and\
  \citenamefont {Serduke}}]{GLOECKNER1972597}%
  \BibitemOpen
  \bibfield  {author} {\bibinfo {author} {\bibfnamefont {D.}~\bibnamefont
  {Gloeckner}}, \bibinfo {author} {\bibfnamefont {M.}~\bibnamefont
  {MacFarlane}}, \bibinfo {author} {\bibfnamefont {R.}~\bibnamefont {Lawson}},
  \ and\ \bibinfo {author} {\bibfnamefont {F.}~\bibnamefont {Serduke}},\ }\href
  {\doibase http://dx.doi.org/10.1016/0370-2693(72)90604-1} {\bibfield
  {journal} {\bibinfo  {journal} {Phys. Lett. B}\ }\textbf {\bibinfo {volume}
  {40}},\ \bibinfo {pages} {597 } (\bibinfo {year} {1972})}\BibitemShut
  {NoStop}%
  \bibitem [{\citenamefont {Schneider}\ \emph {et~al.}(1975)\citenamefont
  {Schneider}, \citenamefont {Gonsior},\ and\ \citenamefont
  {G{\"u}nther}}]{SCHNEIDER1975103}%
  \BibitemOpen
  \bibfield  {author} {\bibinfo {author} {\bibfnamefont {W.}~\bibnamefont
  {Schneider}}, \bibinfo {author} {\bibfnamefont {K.}~\bibnamefont {Gonsior}},
  \ and\ \bibinfo {author} {\bibfnamefont {C.}~\bibnamefont {G{\"u}nther}},\
  }\href@noop {} {\bibfield  {journal} {\bibinfo  {journal} {Nucl. Phys. A}\
  }\textbf {\bibinfo {volume} {249}},\ \bibinfo {pages} {103 } (\bibinfo {year}
  {1975})}\BibitemShut {NoStop}%
  \bibitem [{\citenamefont {Chiang}\ \emph {et~al.}(1980)\citenamefont {Chiang},
  \citenamefont {Wang},\ and\ \citenamefont {Han}}]{0305-4616-6-3-010}%
  \BibitemOpen
  \bibfield  {author} {\bibinfo {author} {\bibfnamefont {H.~C.}\ \bibnamefont
  {Chiang}}, \bibinfo {author} {\bibfnamefont {M.~C.}\ \bibnamefont {Wang}}, \
  and\ \bibinfo {author} {\bibfnamefont {C.~S.}\ \bibnamefont {Han}},\ }\href
  {http://stacks.iop.org/0305-4616/6/i=3/a=010} {\bibfield  {journal} {\bibinfo
   {journal} {J. Phys. G}\ }\textbf {\bibinfo {volume} {6}},\ \bibinfo {pages}
  {345} (\bibinfo {year} {1980})}\BibitemShut {NoStop}%
\bibitem [{\citenamefont {Mach}\ \emph {et~al.}()\citenamefont {Mach} \emph
  {et~al.}}]{doi:10.1142/9789812702401_0042}%
  \BibitemOpen
  \bibfield  {author} {\bibinfo {author} {\bibfnamefont {H.}~\bibnamefont
  {Mach}} \emph {et~al.},\ }\enquote {\bibinfo {title} {Nuclear structure
  studies of exotic nuclei via the strength of e2 transitions; advanced
  time-delayed $\gamma\gamma$ spectroscopy at the extreme},}\ in\ \href@noop {}
  {\emph {\bibinfo {booktitle} { Y. Suzuki, S. Ohya, M. Matsuo, T. Ohtsubo (Eds.), Proc. Int. Symposium: A New Era of Nuclear Structure Physics, Niigata, Japan, 2003, World Scientific, Singapore}}},\
  Chap.~\bibinfo {chapter} {42}, pp.\ \bibinfo {pages} {277--283}\BibitemShut
  {NoStop}%
  \bibitem{mach2017}H. Mach et al.
Phys. Rev. C 95, 014313 (2017).
\end{thebibliography}%

\end{document}